\documentclass{ws-rv9x6}
\usepackage{subfigure}  % to produce side-by-side / subfigures
\usepackage{ws-rv-van}  % numbered citation/references
\usepackage{ws-index}   % to produce multiple indexes

\makeindex
\newcommand{\nn}{\nonumber}

\newcommand{\cM}{ \mathcal{M} }

\begin{document}

\chapter*{Calabi-Yau Three-folds: Poincar\'e Polynomials and Fractals}
%\aindx{Author, F.}                       % author index entry
\author{Anthony Ashmore}
\address{Rudolf Peierls Centre for Theoretical Physics,\\
1 Keble Road, University of Oxford OX13NP, UK; \\
Jadwin Hall, Princeton University, NJ 08544 USA.}
\author[A.~Ashmore \& Y.-H.~He]{Yang-Hui He}
\address{
Department of Mathematics, City University, London,\\
Northampton Square, London EC1V 0HB, UK;\\
School of Physics, NanKai University, Tianjin, 300071, P.R.~China;\\
Merton College, University of Oxford, OX14JD, UK
}

%%%%%%%%%%%%%%%%%%
\begin{abstract}
We study the Poincar\'e polynomials of all known Calabi-Yau three-folds
as constrained polynomials of Littlewood type, thus generalising the well-known investigation into the distribution of the Euler characteristic and Hodge numbers.
We find interesting fractal behaviour in the roots of these polynomials in relation to the existence of isometries, distribution versus typicality, and mirror symmetry.
\end{abstract}
\body

\section{Introduction and Prospects}
The study of Calabi-Yau three-folds has become a vast and important subject.
After almost two decades of explicit construction since that of the quintic hypersurface in complex projective space of dimension four, it still remains an open problem to classify such spaces. This contrasts drastically with Calabi-Yau manifolds in complex dimensions
one and two, of which there are only the two-torus, four-torus and the K3 surface.

Nevertheless, extraordinary progress has been made in cataloguing three and four-folds, giving rise to many new insights.
Of particular note is the work of Kreuzer and Skarke\cite{Kreuzer:2000xy} which produced Calabi-Yau manifolds as hypersurfaces in toric varieties; for three-folds, this amounted to an impressive 473,800,776 explicit examples.
Various other constructions, such as complete intersection three-folds in products of projective spaces and algebraic quotients, have also been fruitful.
Recently, potential interest in particle and string phenomenology has led to the study of three-folds with relatively small Hodge numbers \cite{Candelas:2007ac,Candelas:2008wb,Braun:2010vc,Davies:2011fr,He:2011rs}.

From this vast database, important structures can be observed.
For example, a striking image is found by plotting twice the difference versus the sum of the relevant Hodge numbers $h^{1,1}$ and $h^{2,1}$ of the manifolds, first performed by Candelas et al.\cite{Candelas:1989hd} The resulting symmetry about the vertical axis gives an excellent visualization of mirror symmetry. The paucity of manifolds at the tip of the plot suggests a certain special corner in this landscape \cite{Candelas:2007ac}.
It is therefore natural to beg for more ``experimental'' quantities indicative of perhaps unseen mathematics and physics.

An attempt \cite{He:2010sf}
was initiated to study a generalisation of the Euler characteristic, the Poincar\'e polynomial. 
The complex roots of the Poincar\'e polynomials of known Calabi-Yau spaces were investigated (Newton polynomials of the affine Toric spaces were also studied, though we shall not delve into this in the present work).
The perspective was inspired by recent work on roots of so-called ``constrained'' polynomials -- those with integer coefficients of specific properties.

Historically, constrained polynomials have provided many questions. Littlewood studied such polynomials with coefficients of $\pm1$, now known as Littlewood polynomials\cite{citeulike:9214273}. 
Odlyzko and Poonen
\cite{citeulike:9072118} studied the zeros of similar random polynomials
with coefficients of $0$ and $1$. They found bounds on the zeros
and fractal-like structures in the distribution of roots. Patterns
found by Borwein and J\"orgenson \cite{citeulike:9072177} within plots
of zeros of constrained, random polynomials showed yet more fractal
behaviour near the boundaries of these objects. A program of intense
computational investigation by Christensen \cite{citeulike:9072155},
J\"orgenson \cite{citeulike:9072130} and Derbyshire \cite{citeulike:9214232}
has led to high resolution plots of the zeros of Littlewood polynomials.

\begin{figure}[ht]
\noindent 
\centerline{\psfig{file=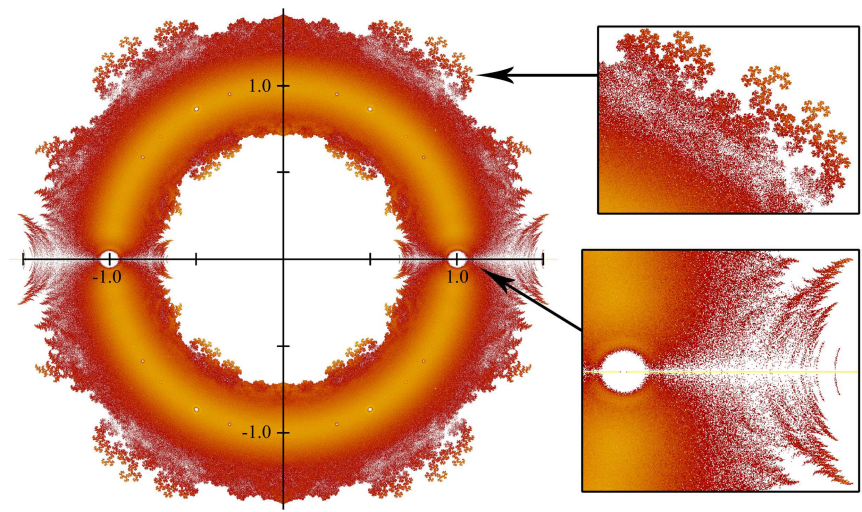,width=1.0\columnwidth}}
\caption{\sf
Plot of complex roots for an order $24$
Littlewood polynomial with $\pm1$ coefficients. Areas which show
self-similarity have been enlarged.
\label{fig:littlewood_roots}
}
\end{figure}

To give the reader of a flavour of these intricacies, in Figure \ref{fig:littlewood_roots} we plot the complex zeros for an order $24$ Littlewood polynomial with random $\pm1$ coefficients.
Self-similar patterns are visible on the boundaries; with this in
mind, we may calculate the Minkowski-Bouligand fractal dimension of the roots using a box counting method afforded by the Matlab${}^{\textregistered}$ package {\sf boxcount} \cite{citeulike:9171409}.
A fractal dimension of $1.90\pm0.08$ was calculated, suggesting a
high degree of statistical self-similarity (the reader is also referred to chaotic behaviour in duality cascades \cite{Franco:2004jz}). 
The emergence of such
delicate features from seemingly simple, mono-variate polynomials
is of great interest; Baez \cite{citeulike:9214157} has recently
posed many questions regarding the nature of the holes and outcrops,
visible in such plots. 

It is the intention of the present note to extend the preceding experiments \cite{He:2010sf} while focusing on the case of compact, smooth, Calabi-Yau three-folds -- we provide analytic insights where possible and delve into the fractal structures where one can.
The organisation and brief summary of the work are as follows.
We introduce the Poincar\'e polynomial for Calabi-Yau manifolds and establish the link with constrained polynomials. Using analytic methods, we present results for self-mirror manifolds and a general solution for all roots of these polynomials, including formulae for the roots using a method which extends
to ninth-order palindromic polynomials.

We then adopt the philosophy of ``experimental'' mathematics and approach the problem from a numerical perspective. The distribution of Hodge numbers for
known Calabi-Yaus and their associated roots are plotted. Delicate
structures, such as the concentration of roots on the unit circle,
are seen, some of which are amenable to analytic explanation. 
Though a comparison with a ``background'' of randomly generated roots shows the distribution does not significantly deviate from the latter, the richness of the structure, both analytic and empirical, suggest that further study should be fruitful.
We have mapped mirror symmetry to a coloured density plot; can certain conformal transformations elucidate these new regions? What statistics can differentiate the true Calabi-Yau nature of a three-fold? Indeed, could the roots hint at ``special'' Calabi-Yau manifolds conducive to a vacuum selection?
These and many more questions await further investigation.

%%%%%%%%%%%%%%%%%%%%%%%%%%%%%%%%%%%%%%%%%%%%%%%%%%%%%%%%%%%
\section{Calabi-Yau three-folds, Hodge numbers and Poincar\'e polynomials}
We begin by gathering some preliminaries in order to set the nomenclature.
The Poincar\'e polynomial, $P_{\mathcal{M}}\left(t\right)$, of a compact, smooth manifold $\cM$ of real dimension $n$ is the
generating function for the Betti numbers of $\cM$:
\begin{equation}\label{P}
P_{\mathcal{M}}\left(t\right)=\sum_{i=0}^{n}b_{i}t^{i} \ .
\end{equation}
Due to the Poincar\'e duality of the Betti numbers, the polynomial
will be palindromic: i.e., the coefficients of $t^{i}$ and $t^{n-i}$
will be equal. Throughout the following discussion, we label the roots of
the equation $P_{\mathcal{M}}\left(t\right)=0$ as $t=\alpha_{i}$.

We note, $P_{\mathcal{M}}\left(-1\right)=\chi$, the Euler characteristic
for $\cM$; it is in this sense that we can think of the polynomial as a generalisation of the important topological quantity $\chi$.
The Poincar\'e-Hopf theorem states that a manifold admits
a vector field without zeros if and only if $\chi=0$, thus the Euler characteristic gives a link to the {\it rank}, or
the number of isometries on a manifold.
In fact, recalling that the rank of $\cM$ is the maximal number of 
everywhere independent, mutually commuting, vector fields thereon, it is an interesting fact that this rank exceeds 1 if and only if $
-1$ is a multiple root of $P_{\cM}(t)$.

For compact, smooth, Calabi-Yau three-folds, \eqref{P} simplifies due to the Hodge diamond structure:
\begin{equation}
{\small
\begin{array}{ccccccccccl}
 &  &  & 1 &  &  & &&&  & b_{0}=1\\
 &  & 0 &  & 0 &  & &&& & b_{1}=0\\
 & 0 &  & h^{1,1} &  & 0 & &&& & b_{2}=h^{1,1}\\
1 &  & h^{2,1} &  & h^{2,1} &  & 1 &&\Rightarrow&& b_{3}=2+2h^{2,1}\\
 & 0 &  & h^{1,1} &  & 0 &  &&&& b_{4}=h^{1,1}\\
 &  & 0 &  & 0 &  &  &&&& b_{5}=0\\
 &  &  & 1 &  &  &  &&&& b_{6}=1 \ ,
\end{array}
}
\end{equation}
giving us the bi-parametric form of the Poincar\'e polynomial as
\begin{equation}
P_{\mathcal{M}}\left(t\right)=1+h^{1,1}t^{2}+\left(2+2h^{2,1}\right)t^{3}+h^{1,1}t^{4}+t^{6} \ .
\label{eq:poincare}
\end{equation}

The strategy is clear: we shall study the space of roots to \eqref{eq:poincare} conglomerated over all known Calabi-Yau three-folds and see what patterns emerge. Indeed, as mentioned before, the plot of $\frac12(P_{\cM}(1) - 4) =  h^{1,1} + h^{2,1}$ drawn vertically against $P_{\cM}(-1) = \chi = 2(h^{1,1} - h^{2,1})$ drawn horizontally has become an iconic image in modern mathematical physics.

\begin{figure}[ht!]
\noindent 
\centerline{\psfig{file=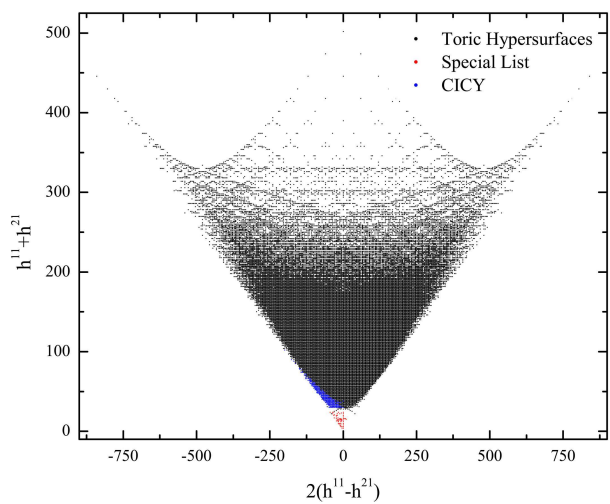,width=0.7\columnwidth}}
\caption{{\sf
\label{fig:hodge}A plot of the Hodge numbers for known Calabi-Yau
three-folds coloured according to the list they are drawn from. The
horizontal axis is $2(h^{1,1}-h^{2,1})=\chi$, the Euler characteristic;
the vertical axis is  $\frac12(P_{\cM}(1) - 4) =  h^{1,1} + h^{2,1}$.
Calabi-Yau three-folds which lie on the vertical line through the origin have
$\chi=0$.
}}
\end{figure}

\begin{figure}
\noindent 
%(a)\includegraphics[width=0.5\columnwidth]{hodge_density}
%(b)\includegraphics[width=0.5\columnwidth]{self_mirror}
\centerline{
(a)\psfig{file=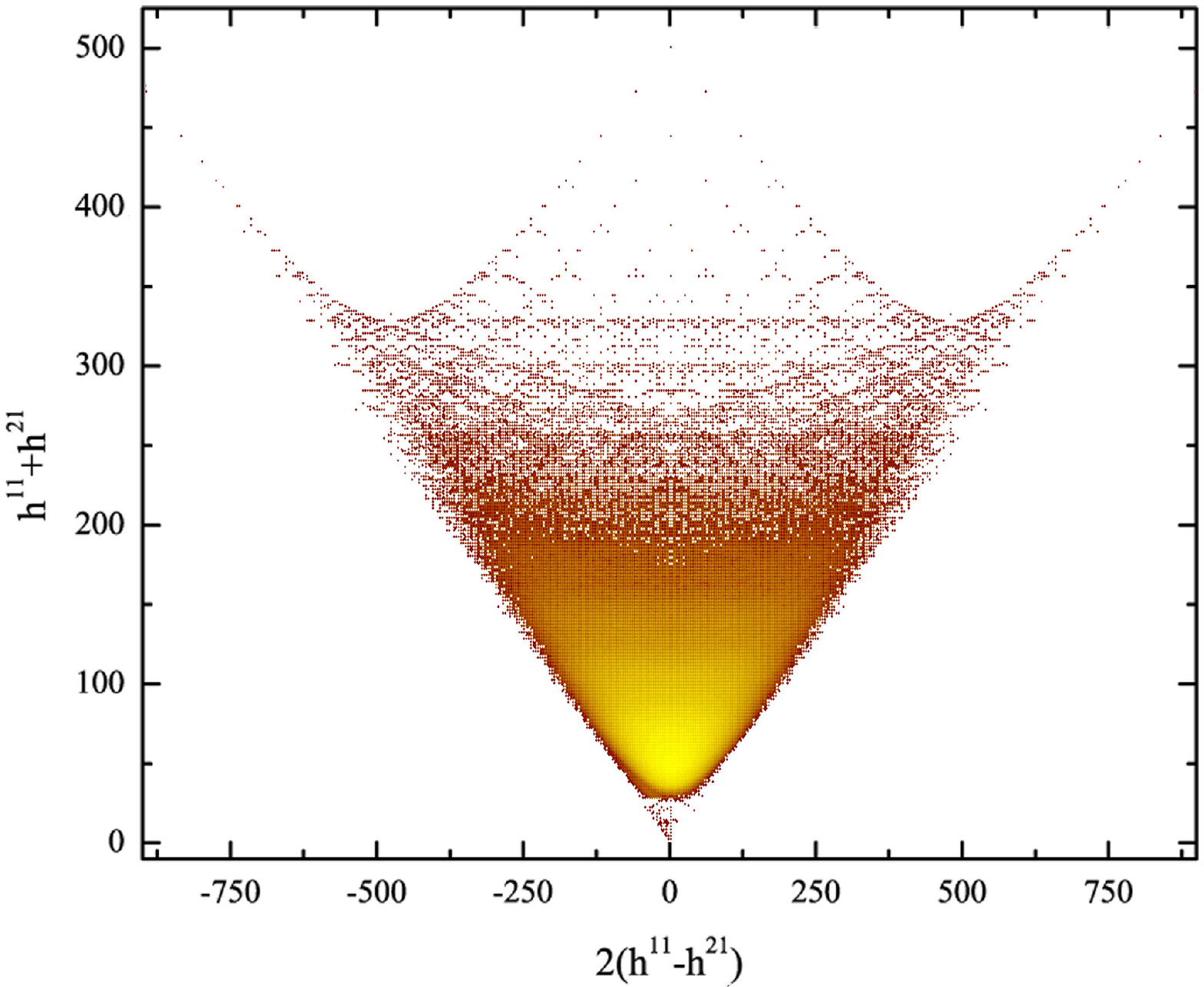,width=0.5\columnwidth}
(b)\psfig{file=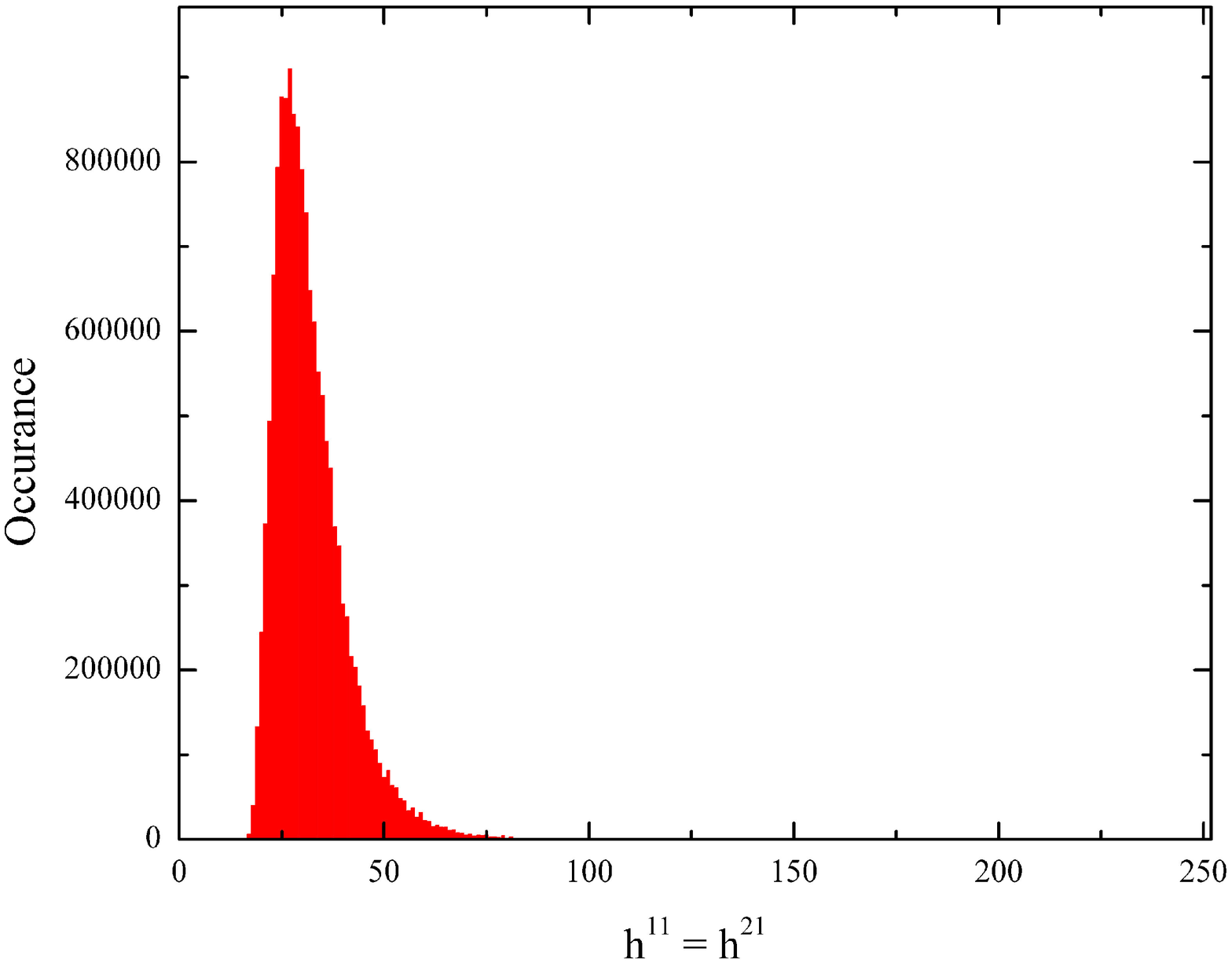,width=0.40\columnwidth}
}
\caption{{\sf
\label{fig:hodge_density}
(a)
A plot of the density of Hodge numbers of all known Calabi-Yau three-folds.
As standard, $\frac12(P_{\cM}(1) - 4) =  h^{1,1} + h^{2,1}$ is the ordinate while $P_{\cM}(-1) = \chi = 2(h^{1,1} - h^{2,1})$ is the abscissa,
In colour rendering, red indicates low density while yellow is high density.
(b)
A histogram plot of the self-mirror three-folds.
There is a dramatic peak around $(h^{1,1},h^{2,1})=(27,27)$ in both (a) and (b).
}}
\end{figure}

For reference, we reproduce this plot of the Hodge numbers for known Calabi-Yau three-folds in Figure \ref{fig:hodge}.
There is an obvious symmetry about the vertical
axis at $2\left(h^{1,1}-h^{2,1}\right)=\chi=0$, this is the best experimental 
evidence for {\it mirror symmetry}. 
We note the manifolds with unpaired mirrors towards the bottom of the plot. 
These are from complete intersection manifolds in products of projective spaces, ironically the earliest of the databases of Calabi-Yau three-folds constructed \cite{Candelas:1987kf}.
The mirrors for these spaces are yet to be discovered due to the lack of a
systematic construction; explicit construction requires possibly complicated quotienting.

The pair with the largest Hodge number is
$\left(491,11\right)$; a Calabi-Yau three-fold with $h^{1,1}$ or
$h^{2,1}$ exceeding $491$ is yet to be found. Yau conjectured that
there are a finite number of topologically distinct Calabi-Yau manifolds
in each dimension. This conjecture is still
open, but is spurred on by the apparent lack of geometries with larger
Hodge numbers.

Let us refine the famous plot by giving it an extra dimension.
Distinct Calabi-Yau three-folds may have identical Hodge numbers -- this can result in enormous degeneracy for each $(h^{1,1}, h^{2,1})$ pair.
The density of the Hodge numbers is shown in Figure \ref{fig:hodge_density} (a) with yellow signifying greater degeneracy.
There is an approximate uniform density towards the centre of the
plot, with lower densities in the feathered regions for large $h^{1,1}+h^{2,1}$. A high concentration of geometries can be seen near the lower tip
of the triangle. There is a total of 15,067,026 self-mirror manifolds, lying on the line of symmetry in the plot with $h^{1,1} = h^{2,1}$, giving 149 distinct Hodge number pairs.
For reference, we plot the frequency distribution of such manifolds in Figure \ref{fig:hodge_density} (b) and see the sharp peak at 27. In fact, $(27,27)$ dominates the space of known Calabi-Yau three-folds with 910,113 such manifolds. This is followed by 877,191 at (25,25) and 875,275 at (26,26).
Thus, one could consider these as {\it typical} Calabi-Yau three-folds in the landscape of possibilities.

%%%%%%%%%
\subsection{Calabi-Yau Roots: Analytic Results}
Having presented the data of known three-folds, we now move on to address the space of roots of the Poincar\'e polynomials.
First, we will examine \eqref{eq:poincare} analytically to filter out ``background'' effects from generic sextic behaviour. Next, we will turn to explicit data and see features specific to the Calabi-Yau data.

A general sextic polynomial is not solvable by algebraic methods. We first examine Calabi-Yau three-folds with zero Euler
characteristic, corresponding to self-mirror manifolds.
In these cases, the polynomial may be reduced to a lower order and
solved explicitly.
The palindromic constraint is shown
to manifest itself as the appearance of roots in inverse pairs. Finally,
a complete solution for the general form of $P_{\mathcal{M}}\left(t\right)$
is presented and shown to be applicable up to ninth-order polynomials
with similar constraints.

Our polynomial has only integer coefficients, implying that possible
integer roots lie in the set of exact divisors of the coefficient
of $t^{0}$, viz., $b_{0}$. However, for all our spaces, due to connectedness,
$b_{0}=1$. Hence, we know that the only integer roots of
this polynomial are $\pm1$. Given that all the coefficients $b_{i}\geq 0$, 
we can eliminate $1$ as a root, 
leaving $-1$ as the only \emph{integer} root of the
Poincar\'e polynomial. Given that the coefficients are positive, we expect
only negative and complex roots. Finally, as all coefficients are real, 
if $\alpha_{i}$ is a complex root of the polynomial
it follows that $\alpha_{i}^{*}$ is also a root.

%%%
\subsubsection{Zero Euler Characteristic}
Let us first try a natural simplification. Evaluating the Poincar\'e polynomial at $t=-1$ gives the Euler characteristic for the space; if $-1$ is actually a root, we have a manifold with $\chi=0$. This occurs for self-mirror manifolds, $h^{1,1}=h^{2,1}$. For this special case, we may factor this root out in an effort to reduce our sextic
to a lower order polynomial:
\begin{eqnarray}
(1+t)^{2}(1-2t+(3+h^{1,1})t^{2}-2t^{3}+t^{4}) =  0 \ .
\end{eqnarray}

We see that $t=-1$ is (at least) a double root. The resulting quartic equation \emph{does} have a general solution: 
{\small
\[\begin{array}{l}
\{ \frac{1}{2}+\frac{1}{2}\sqrt{-3-2i\sqrt{h^{1,1}}-h^{1,1}}-\frac{i\sqrt{h^{1,1}}}{2} \ ,
\frac{1}{2}+\frac{1}{2}\sqrt{-3+2i\sqrt{h^{1,1}}-h^{1,1}}+\frac{i\sqrt{h^{1,1}}}{2},\nn \\
\frac{1}{2}-\frac{1}{2}\sqrt{-3-2i\sqrt{h^{1,1}}-h^{1,1}}-\frac{i\sqrt{h^{1,1}}}{2} \ ,
\frac{1}{2}-\frac{1}{2}\sqrt{-3+2i\sqrt{h^{1,1}}-h^{1,1}}+\frac{i\sqrt{h^{1,1}}}{2} \} \ .
\end{array}
\]
}

%%%
%\subsubsection{Right Half of Unit Circle}\label{sub:unit_circle}
%We look for solutions which lie on the unit circle -- $t=e^{i\theta}$.
%Solutions to \eqref{eq:poincare} must then satisfy

%\begin{equation}
%\frac{(2+2h^{2,1})}{8}+\frac{(2h^{1,1}-6)}{8}\cos\theta+\cos^{3}\theta=0 \ .
%\end{equation}

%For solutions on the unit circle with zero or positive real part,
%$\cos\theta$ lies in the interval $\mbox{\ensuremath{\left[0,1\right]}}$.
%Given that $h^{1,1}\geq1,\, h^{2,1}\geq0$ and are integers, it is
%simple to see that no solutions exist. Thus there are no solutions
%on the unit circle for $\Re\textnormal{e}(t)\geq0$.

%%%
\subsubsection{Palindromic polynomials}
Now let us comment on general solutions.
Of course, polynomials of degree five or greater evade general algebraic solutions, usually forcing one to resort to numerical calculations. 
However, the generating polynomials arising from our Calabi-Yau spaces are naturally palindromic, allowing us to make some progress.

A simple substitution shows that the polynomial is unchanged by $t\rightarrow1/t$ (this is usually referred to as the polynomial being ``self-reciprocal''; for the appearance and relevance of palindromic polynomials in Hilbert series analyses of Calabi-Yau geometries \cite{Forcella:2008bb}, the reader is referred to Section 2 of cit.~ibid.).
Using this, we may write the polynomial in a more suggestive form by factoring it in terms of its roots.

\begin{eqnarray}
(\frac{1}{t}-\alpha_{1})(\frac{1}{t}-\alpha_{2})(\frac{1}{t}-\alpha_{3})(\frac{1}{t}-\alpha_{4})(\frac{1}{t}-\alpha_{5})(\frac{1}{t}-\alpha_{6}) & = & 0 \nn \\
(t-\alpha_{1})(t-\alpha_{2})(t-\alpha_{3})(t-\alpha_{4})(t-\alpha_{5})(t-\alpha_{6}) & = &
 0 \ .
\end{eqnarray}

For this to be true, it must hold that if $\alpha_{i}$ is a root,
$1/\alpha_{i}$ is also a root. This sixth-order polynomial has only
\emph{three} independent roots. Without loss of generality, we may
identify $\alpha_{4}$ with $\frac{1}{\alpha_{1}}$ etc. We conclude, therefore, that the roots to our self-reciprocal polynomials appear in inverse pairs and that such a sextic polynomial has only three independent roots. It is interesting
to ask if this allows an explicit algebraic solution; can the polynomial
be re-expressed as a solvable cubic? We define the variable $\xi=t+\frac{1}{t}$
and consider the following for even-order\footnote{Substitution of $t=-1$ into an odd self-reciprocal polynomial shows
that this is always a root. This may be factored out to give an even order
polynomial.} self-reciprocal polynomials:
\begin{eqnarray}
P_{\mathcal{M}}\left(t\right) =  \sum_{i=0}^{n}a_{i}t^{i}\enskip\textnormal{where }a_{i}=a_{n-i},\quad Q_{\mathcal{M}}\left(\xi\right)  =  \sum_{j=0}^{\frac{n}{2}}b_{j}\xi^{j} \ .
\end{eqnarray}

We now assert that, given the original polynomial, we can always find
$b_{j}$ such that 
\begin{equation}
t^{\frac{n}{2}}Q_{\mathcal{M}}\left(\xi\right)=P_{\mathcal{M}}\left(t\right) \ .\end{equation}
An explicit proof of this is given by Ahmadi and Vega \cite{citeulike:9071308}.
Briefly, the argument goes as follows.
We have
\begin{eqnarray*}
P_{\mathcal{M}}\left(t\right) =  \sum_{i=0}^{n}a_{i}t^{i} =  t^{\frac{n}{2}}\left\{ a_{\frac{n}{2}}+\sum_{j=1}^{\frac{n}{2}}a_{\frac{n}{2}-j}\left[t^{j}+t^{-j}\right]\right\} \ .
\end{eqnarray*}
Each term in the square brackets
may be rewritten as a function in powers of $\xi=t+\frac{1}{t}$. For example, for $j=4$,
\begin{eqnarray*}
t^{4}+t^{-4} & = & \left(t+t^{-1}\right)^{4}-4\left(t^{2}+t^{-2}\right)-6\\
\textnormal{while}\quad t^{2}+t^{-2} & = & \left(t+t^{-1}\right)^{2}-2\\
\textnormal{giving\quad}t^{4}+t^{-4} & = & \left(t+t^{-1}\right)^{4}-4\left(t+t^{-1}\right)^{2}+2 \ .
\end{eqnarray*}
By induction, any expression of the form $t^{j}+t^{-j}$ may be reduced
to a function of $\left(t+t^{-1}\right)$ only. Hence, the terms inside
the braces may recast as a function of $\left(t+t^{-1}\right)$ too,
giving an $\frac{n}{2}^{\textnormal{th}}$ order function.

The polynomial $Q_{\mathcal{M}}\left(\xi\right)$ is of order $\frac{n}{2}$,
its roots may be solved for explicitly when $n\leq8$. Given that zero
is not a root and that the roots appear in reciprocal pairs, solutions
to $Q_{\mathcal{M}}\left(\xi\right)=0$ will give the roots of $P_{\mathcal{M}}\left(t\right)$.

We examine this in the context of the polynomials we have been considering by expanding $Q_\mathcal{M}\left(\xi\right)$ in $t$ for $\frac{n}{2}=3$. Comparing this with the general form for the Poincar\'e polynomial,
we see this cubic equation may be written as
\begin{equation}
\xi^{3}+(h^{1,1}-3)\xi+(2+2h^{2,1})=0 \ .
\end{equation}
We can solve this cubic explicitly, giving three values for $\xi$. The relation between $\xi$ and
$t$ may then be inverted to solve for $t$. In this way, it is possible to explicitly solve for all roots of our
palindromic polynomial. The six explicit solutions are:
\begin{eqnarray}\label{sol}
t & = & \frac{1}{2}\left(\xi \pm \sqrt{\xi^{2}-4}\right) \ ; \qquad
\mbox{with }\\ \nn
\xi & = & \frac{(3-h^{1,1}) \omega + y^2}{\omega^2 y} \ , 
   \frac{(1+i \sqrt{3})\omega (h^{1,1}-3) + 
     i (i+\sqrt{3})y^2}{2\omega^2\ y} \ , \\
  & &   -\frac{(1+i\sqrt{3})y^2 + (3 - h^{1,1}(1-\sqrt{3}i)}{2\omega^2\ y} \ ,
\\ \nn
\end{eqnarray}
where $\omega = \sqrt[3]{3}$ is the primitive cubic root of 3 and
$y = \sqrt{3 h^{1,1} ((h^{1,1}-9) h^{1,1}+27)+81 h^{2,1} (h^{2,1}+2)}-9 h^{2,1}-9$.

%%%%%%%%%%%%%%%%%%%%%%%%%
\subsection{Calabi-Yau Roots: Numerics}
We now consider actual data and approach the problem from a numerical
perspective. We plot the zeros of various random, constrained polynomials
with coefficients chosen to aid the analysis of roots from Calabi-Yau
spaces. The $38,059$ Hodge number pairs are used as coefficients of
the Poincar\'e polynomial and the zeros are plotted. The density of zeros and the analytic solutions from
which they arise are also shown. Delicate structures are seen, such
as the concentration of roots on the unit circle. This Calabi-Yau
data is then compared with the background of randomly generated roots.

%%%%%%%%%%%%%
\subsubsection{Random Constrained Polynomials}

\begin{figure}[ht!]
%(a) \includegraphics[width=0.4\columnwidth]{rand_no_linear_small}
%(b) \includegraphics[width=0.4\columnwidth]{rand_pal_no_linear_small}
\centerline{
(a)\psfig{file=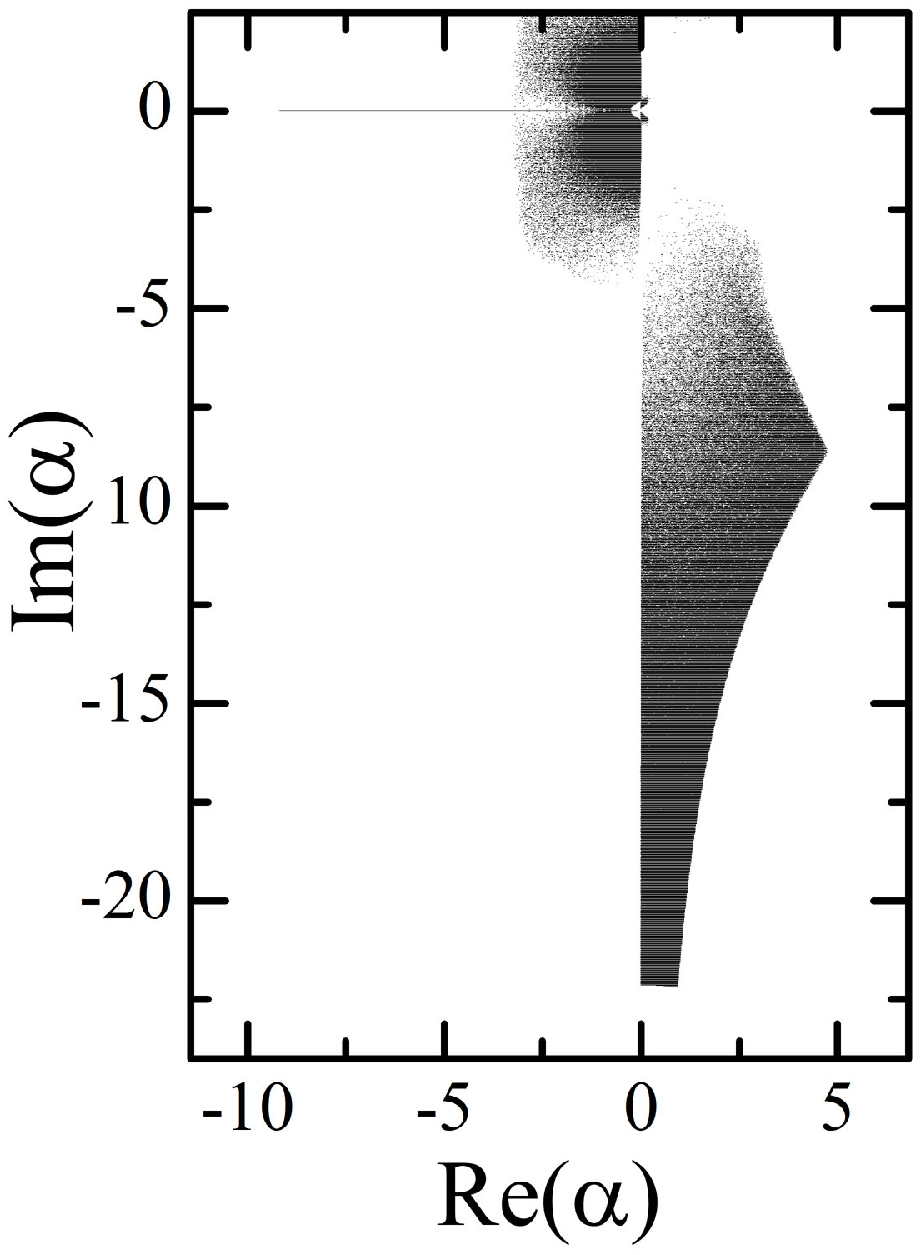,width=0.45\columnwidth}
(b)\psfig{file=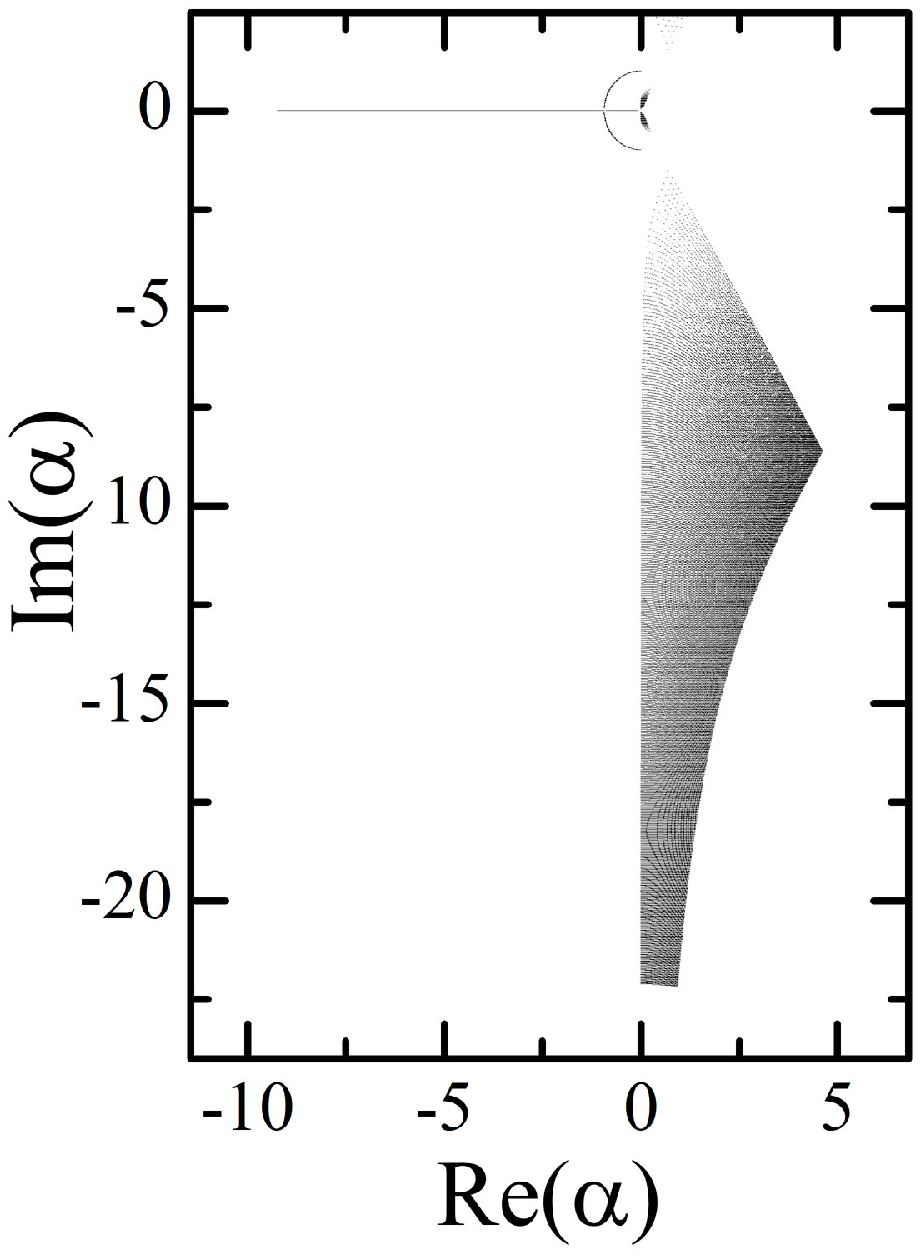,width=0.45\columnwidth}
}
\caption{{\sf
\label{fig:rand_sextic_no_linear}
(a) A plot of the roots for a sextic
polynomial with vanishing linear and quintic terms of the form $P=1+at^{2}+(2+2b)t^{3}+ct^{4}+t^{6}$.
$a$, $b$, $c$ take random values from $\left[1,491\right]$.
(b) A subset of part (a) but using only palindromic polynomials.}
}
\end{figure}

In the hope of finding an underlying pattern to our Calabi-Yau data,
it is prudent to first examine similar polynomials with
randomly generated coefficients. 
In part (a) of Figure \ref{fig:rand_sextic_no_linear}, we plot the
zeros of a sextic with vanishing linear and quintic terms.\footnote{Only the lower half plane is shown. By complex conjugation, there
is a symmetry about the real axis.} The coefficients of $t^{0}$ and $t^{6}$ are set to unity. The
remaining coefficients are chosen to give the same form as the polynomials
which arise from Calabi-Yau spaces \emph{without} requiring $a_{2}=a_{4}$,
i.e. we have not imposed palindromicity. The coefficients are randomly
chosen from the range $\left[1,491\right]$; this allows a direct
comparison with real Hodge numbers which lie within the same range.
The roots are localised near the origin and in two lobes extending
into the upper and lower half planes.

We restrict further to palindromic sextics and plot the zeros in part (b) of
Figure \ref{fig:rand_sextic_no_linear}. There is a concentration
of zeros on the left half of the unit circle. There is a more distinct boundary to the
roots than the previous case and a wedge, on the right half plane with opening angle  $\frac{2\pi}{3}$, devoid
of any zeros. Allowing the coefficients to take values greater than $491$ extends
the lobes in the upper and lower planes while roots lying on the real
axis spread to larger negative values.

\begin{figure}[ht!]
\noindent 
\centering{\psfig{file=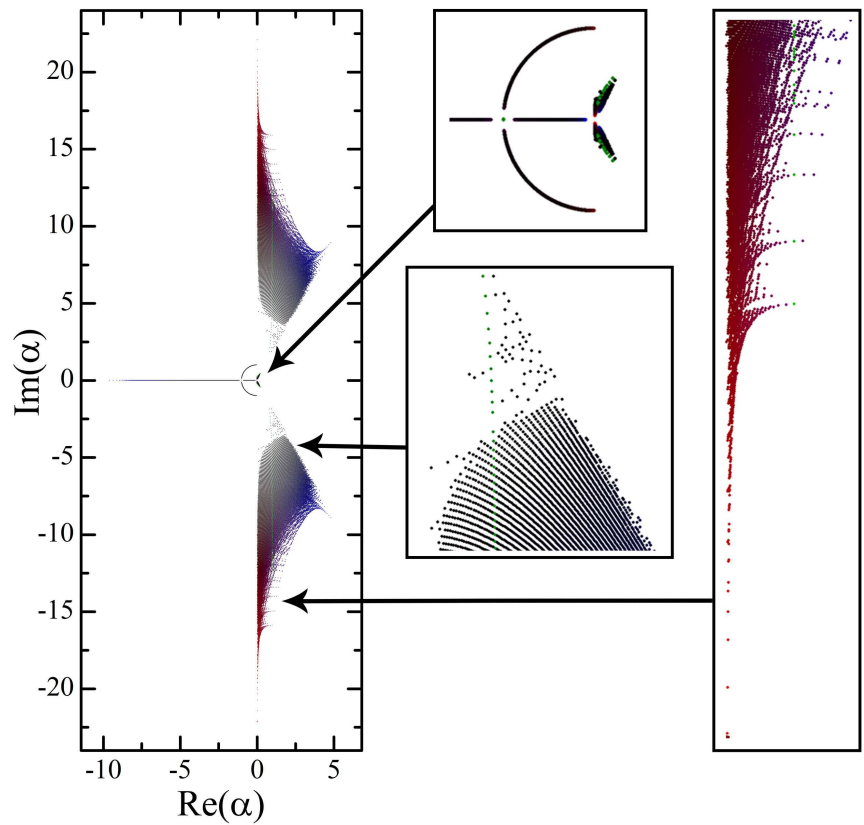,width=1.0\columnwidth}}
\caption{{\sf
\label{fig:roots}A plot of the roots for known Calabi-Yau three-folds
in the complex plane. Red corresponds to large values of $h^{1,1}$,
blue to large $h^{2,1}$ and green for $h^{1,1}=h^{2,1}$. Areas of
interest are shown enlarged.
}}
\end{figure}

The list of known Calabi-Yau Hodge numbers was imported into Matlab
and used as the coefficients of Poincar\'e polynomials. The roots, $\alpha_{i}$,
are plotted on the complex plane in Figure \ref{fig:roots}. The points
are coloured according to the values of the Hodge numbers associated
with them: bright red indicates large values of $h^{1,1}$, blue are
large $h^{2,1}$. Roots which correspond to self-mirror manifolds
with $h^{1,1}=h^{2,1}$ are coloured green. Areas of interest have
been enlarged to display the intricate patterns present. 

Roots on the unit circle and the negative
real axis approach, but do not reach, $-1$, unless they are self-mirror
manifolds. There are also a pair of outcrops from the origin which
appear to be isolated from the other roots. The integer nature of
the coefficients leads to a feathering effect on the boundaries, similar
to that seen for Littlewood polynomials in Figure \ref{fig:littlewood_roots}.

Let us now compare the Calabi-Yau data with the comparable random
background in Figure \ref{fig:rand_sextic_no_linear} (b). We notice
\emph{no significant differences} between the two plots -- this suggests
that such roots cannot be used to classify Calabi-Yau spaces. Mirror
symmetry acts by interchanging red and blue points. There is no obvious
symmetry present in the roots which might explain how mirror manifolds
are related to each other. Various conformal mappings have been tested
with the hope of making mirror symmetry manifest in the roots: no
mapping was found which gave the desired effect.

\begin{figure}[ht!]
\noindent 
%(a)\includegraphics[width=0.4\columnwidth]{density_plot_small}
%(b)\includegraphics[width=0.4\columnwidth]{analytic_roots}
\centerline{
(a)\psfig{file=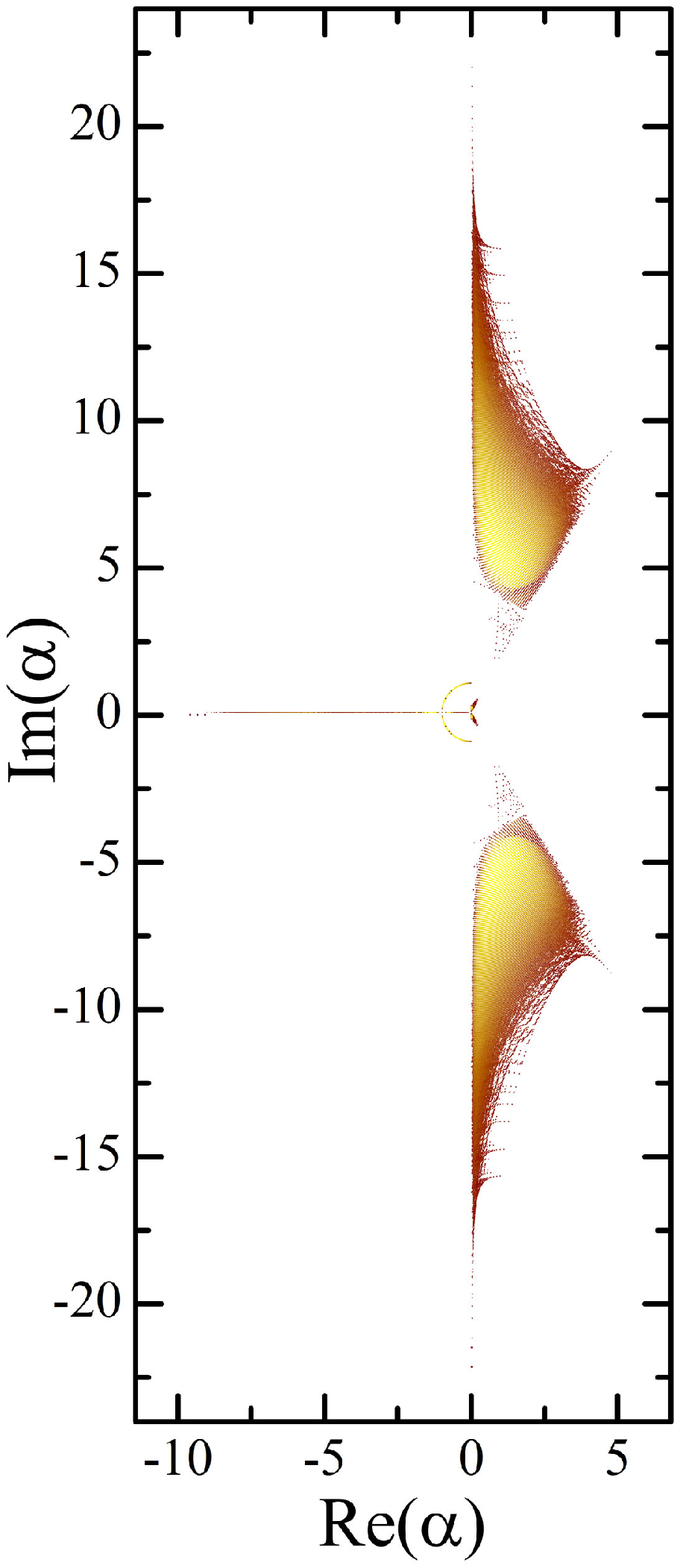,width=0.42\columnwidth}
(b)\psfig{file=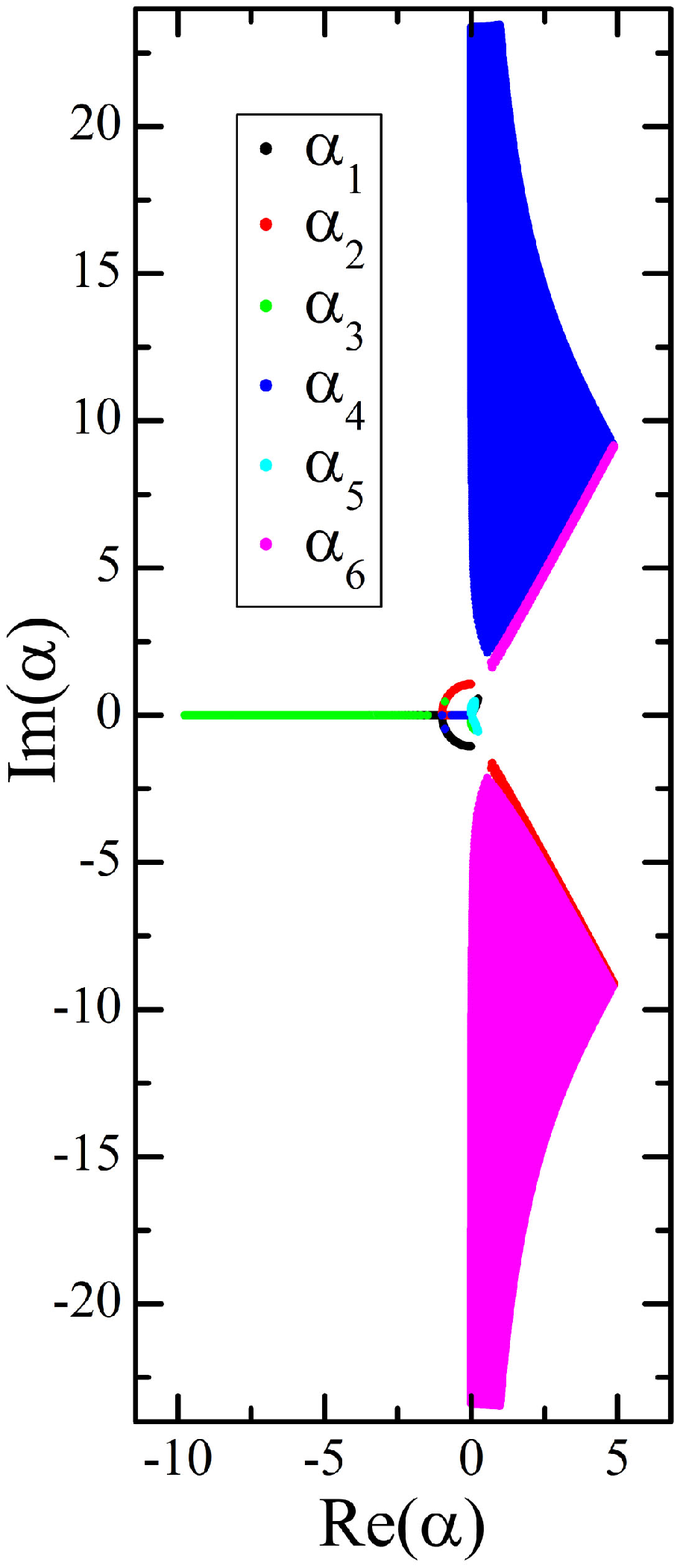,width=0.42\columnwidth}
}
\caption{{\sf
\label{fig:roots_density}
(a) A plot of the density of roots for Calabi-Yau
three-folds in the complex plane. Red indicates low density while
yellow signifies high density.
(b) A plot of the analytic roots for a sextic
palindromic polynomial with vanishing linear and quintic terms of
the form $P=1+at^{2}+(2+2b)t^{3}+at^{4}+t^{6}$ where $a$, $b$ take
random values in the range {[}$1$,$491${]}.
}}
\end{figure}

The density of the roots are plotted in part (a) of 
Figure \ref{fig:roots_density}.
There is a clustering of points both on the left half of the unit
circle and the negative real axis; this behaviour stems from the palindromic
nature of the generating polynomial and is seen for the randomly generated
sextics too.

Using the explicit formulae from \eqref{sol},
in part (b) of Figure  \ref{fig:roots_density}
we plot the roots on the complex plane for $h^{1,1},\, h^{2,1}\in\left[1,491\right]$.
Points are coloured according to the solution used to calculate the
root. It is hoped that this may prove useful in future work when studying
the effect of mirror symmetry on the zeros of the Poincar\'e polynomials.

\section*{Acknowledgments}
{\it Ad piam memoriam Maximiliani Kreuzer hoc opusculum dedicatum est.}
AA is grateful to Jesus College, the University of Oxford and Princeton University for their support.
YHH would like to thank the Science and Technology Facilities
Council, UK, for an Advanced Fellowship and grant ST/J00037X/1, the Chinese Ministry of Education, for a Chang-Jiang Chair Professorship at NanKai University, the US NSF for grant CCF-1048082,
as well as City University, London and Merton College, Oxford, for their enduring support.

%%%%%%%%%%%%%%%%%%%%%%%%%%%%%%%%%%%%%%%%%%%%%%%%%%

\end{document}